\documentclass[journal=jctcce,manuscript=article]{achemso}
\usepackage{siunitx}
\usepackage{amsfonts}
\usepackage[pdftex]{thumbpdf} 
\usepackage[bookmarks = true, 
            bookmarksnumbered = true, 
            pdfpagemode = None, 
            pdfstartview = FitH, 
            pdfpagelayout = SinglePage, 
            colorlinks = true, 
            urlcolor = magenta, 
            linkcolor = black, 
           ]{hyperref} 

\hypersetup{ 
            pdfauthor = {pcc},
            pdftitle = {guoshu},
            pdfsubject = {GDRI},
            pdfkeywords = {ENMFCI},
            pdfcreator = {PDFLaTex},
            pdfproducer = {PDFLaTex}}
\usepackage[latin1]{inputenc}
\usepackage{amsmath}
\usepackage{amsfonts}
\usepackage{amssymb}
\usepackage{textcomp}
\usepackage{xtab}
\usepackage[english]{babel}

\newcommand{\sch}{Schr\"{o}dinger\ } 

\newcommand{\JCPformat}[4]{{#1} {\bf #2}, {#3} ({#4}).}

\newcommand{\Ref}[4]{\JCPformat{#1}{#2}{#3}{#4}}

\newcommand{\jcp}[3]{\Ref{J. Chem. Phys}{#1}{#2}{#3}}

\newcommand{\tca}[3]{\Ref{Theor. Chim. Acta.}{#1}{#2}{#3}}
\newcommand{\jmathchem}[3]{\Ref{J. Math. Chem.}{#1}{#2}{#3}}

\newcommand{\molphys}[3]{\Ref{Mol. Phys.}{#1}{#2}{#3}}

\newcommand{\physrev}[3]{\Ref{Phys. Rev.}{#1}{#2}{#3}}

\author{Patrick Cassam-Chena\"\i}
\email{cassam@unice.fr}
\author{Gilles Lebeau}
\affiliation[University of Nice Sophia Antipolis]{Universit\'e C\^ote d'Azur, LJAD, UMR 7351, 06100 Nice, France}

\title{Smeared Coulomb potential orbitals for the electron-nucleus mean field configuration interaction method}

\begin{document}
\selectlanguage{english}

\begin{abstract}

We propose to use the eigenfunctions of a one-electron model Hamiltonian to perform electron-nucleus mean field configuration interaction (EN-MFCI) calculations. The potential energy of our model Hamiltonian corresponds to the Coulomb potential of an infinite wire with charge $Z$ distributed according to a Gaussian function. The time independent \sch equation for this Hamiltonian is solved perturbationally in the limit of small amplitude vibration (Gaussian function width close to zero).

\end{abstract}

\newpage
\section{Introduction}
This paper is dedicated to Prof. Graham Chandler, whose famous ``Mclean and Chandler basis sets'' have proved extremely useful to the quantum chemistry community. Some twenty years ago, we  optimized Gaussian basis sets for molecular fragments together with D. Jayatilaka and G. S. Chandler \cite{Cassam98-jchimphys}. However, this endeavour was suspended due to technical difficulties raised by electronegative atoms. In the present article, we come back to, perhaps, a more original approach to basis functions, where the latter are not selected on the ground of their technical advantages, as was the case initially for Gaussian-type orbitals (GTO) \cite{Boys70}, but because they are eigenfunctions of a model Hamiltonian and therefore have some physical relevance. It is hoped that this property can be taken advantage of in basis set truncation.\\

The model Hamiltonian we will consider is a generalization of the hydrogenoid atom Coulomb Hamiltonian.
So, our new orbitals will be part of the exponential-type orbitals (ETO) \cite{Guseinov2014} family, as the hydrogenoid atom eigenfunctions, which constitute an asymptotic limit. Slater-type orbitals (STO) \cite{Zener30,Slater30} is another type of ETO related to hydrogenoid orbitals (HO): they can be seen as ``uncontracted'' HO, that is to say, as HO with the Laguerre polynomial prefactor replaced by a simple monomial one.  
In contrast to GTO, the difficulty of computing multicenter integrals with STO has led to the introduction of more ETO family members such as  Bessel-type orbitals (BTO) \cite{Filter78} or Coulomb-Sturmian orbitals (CSO) \cite{Shull59}.  The techniques developed for the latter \cite{Avery2013,Avery2015,Avery2018} will be equally relevant for integrals involving our deformed hydrogenoid orbitals (DHO).\\

It is our take that DHO will be particularly useful for the recently developed electron-nucleus mean field configuration interaction (EN-MFCI) method \cite{Cassam15-pra}.  The EN-MFCI method affords one to obtain in a single calculation, the electronic and vibrational energy levels of a molecule, without making the ``Born-Oppenheimer'' (BO) approximation \cite{Cassam15-pra,Cassam17-tca}.  In contrast, the traditional methods of Quantum Chemistry are set in the frame of this approximation.  They describe electronic clouds of fixed nuclear configurations and make use of orbital basis sets centered on nuclear positions.  The latter basis sets are not appropriate for EN-MFCI calculations and their discrepancies have been bypassed so far, only by adding off-centered orbitals.  However, these additional functions introduce linear dependencies with the initial orbitals and span virtual molecular orbitals of little relevance for the description of low energy wave functions of the molecule.  So, it appears important to develop new orbital basis sets for the EN-MFCI method, able to describe the electron cloud of oscillating nuclei in a molecular system.\\

We propose to use the eigenfunctions of a  one-electron model  Hamiltonian:\\
$H=-\frac{\vartriangle}{2\mu}+V(\vec{r})$, with a  potential of the form: \\
\begin{equation}
V(\vec{r})=-Z\sqrt{\frac{a}{\pi}}\int\limits_{-\infty}^{+\infty} \frac{exp[-az_0^2]}{\Arrowvert\vec{r}-\vec{r_{z_0}}\Arrowvert} dz_0,
\end{equation}
where, $Z\in\mathbb{N}$, $a\in\mathbb{R}^+$, $\vec{r}=(x,y,z)$ and $\vec{r_{z_0}}=(0,0,z_0)$ in Cartesian coordinates.
When $\vec{r}$ is expressed in  cylindrical coordinates, $\vec{r}=(\rho,\phi,z)$, the potential depends only upon $\rho$ and $z$,
\begin{equation}
V(\rho,z)=-Z\sqrt{\frac{a}{\pi}}\int\limits_{-\infty}^{+\infty}\frac{exp[-az_0^2]}{\sqrt{\rho^2+(z-z_0)^2}} dz_0.
\end{equation}
This potential corresponds to the Coulomb  potential of an infinite wire with charge $Z$ distributed according to a Gaussian function.  
In the limiting case of a Gaussian function sharply peaking at the origin ($a\rightarrow +\infty$, Dirac distribution limit), the system will tends towards a point-charge Z concentrated at the origin and the hydrogenoid atom eigenfunctions will be recovered.\\
However, by taking a Gaussian width parameter of the order of magnitude of a nucleus vibration amplitude, we will get basis functions corresponding to a Coulomb potential convoluted by a nuclear, vibrational, harmonic motion, that we may think particularly appropriate for  EN-MFCI calculations.\\
Unfortunately, the \sch equation for this potential is hard to solve because the whole $z$-axis is singular. So, we will restrict ourselves to the $a\rightarrow\infty$ asymptotic limit, and expand the potential $V(\rho,z)$ as
\begin{small}\begin{eqnarray}
&V(\rho,z)=\frac{-Z\sqrt{a}}{\pi}\int\limits_{0}^{+\infty}\frac{d\lambda}{\sqrt{\lambda}}\int\limits_{-\infty}^{+\infty}dz_0\ exp[-az_0^2-\lambda(\rho^2+(z-z_0)^2)]& \nonumber\\
&=-Z\sqrt{\frac{a}{\pi}}\int\limits_{0}^{+\infty}\frac{d\lambda}{\sqrt{\lambda(a+\lambda)}} exp[-\lambda(\rho^2+z^2)]exp[\frac{\lambda^2}{a+\lambda}z^2]&\nonumber\\
&=\frac{-Z}{\sqrt{\pi}}\int\limits_{0}^{+\infty}d\lambda\left(\frac{1}{\sqrt{\lambda }}+\frac{-\frac{\lambda }{2}+z^2 \lambda ^2}{ \sqrt{\lambda
} a}+\frac{\lambda ^{3/2} \left(3-12 z^2 \lambda +4 z^4 \lambda ^2\right)}{8  a^2}+\frac{\lambda ^{5/2} \left(-15+90 z^2 \lambda -60 z^4
\lambda ^2+8 z^6 \lambda ^3\right)}{48 a^3}+o\left(\frac{1}{a}\right)^{\frac{7}{2}}\right) exp[-\lambda(\rho^2+z^2)]&\nonumber\\
\end{eqnarray}\end{small}
\noindent
Swaping the limits and setting $r=\sqrt{\rho^2+z^2}$, $\rho=r\times sin(\theta)$, $z=r\times cos(\theta)$, with $\theta\in\left[0,\pi\right]$, the potential becomes  
\begin{small}\begin{eqnarray}
&V(r,\theta)=-\frac{Z}{r}+\frac{\left(1-3 cos(\theta)^2\right) Z}{4 r^3 a}-\frac{3 \left(\left(3 -30  cos(\theta)^2+35 cos(\theta)^4\right) Z\right)}{32 r^5 a^2}+\frac{15
\left(5 -105  cos(\theta)^2+315  cos(\theta)^4-231 cos(\theta)^6\right) Z}{128 r^{7} a^3}+O\left[\frac{1}{a}\right]^4,&\nonumber\\
\end{eqnarray}\end{small}
where we recognize the hydrogenoid atom potential in the zero$^{th}$-order term. 
\begin{equation}
V^{(0)}(r,\theta)=-\frac{Z}{r}.
\end{equation}
Now, at any order, the singularity is located at the single point $r=0$. In the following, we will only consider the $V(r,\theta)$ potential truncated at first order,
\begin{equation}
V_{1}(r,\theta):=-\frac{Z}{r}+\frac{\left(1-3 cos(\theta)^2\right) Z}{4 r^3 a}.
\end{equation}
Since this potential is not bounded from below, we will just apply Rayleigh-\sch perturbation theory to first order,
to get corrections with respect to the hydrogenoid atom eigenstates.

\section{Perturbationally corrected eigenstates}
So, we consider the following Hamiltonian in spherical coordinates and atomic units
\begin{equation}
H=\frac{-1}{2\mu} \left(\frac{1}{r^2}\frac{\partial}{\partial r} r^2\frac{\partial}{\partial r} + \frac{1}{r^2 sin(\theta)}\frac{\partial}{\partial \theta} sin(\theta) \frac{\partial}{\partial \theta}+ \frac{1}{r^2 sin(\theta)^2}\frac{\partial^2}{\partial\phi^2} \right) + V_{1}(r,\theta),
\end{equation}
defined on the Hilbert space of square integrable functions whose scalar product is expressed as,
\begin{equation}
 \langle\psi_1|\psi_2\rangle:=\int\limits_{0}^{+\infty}r^2 dr\int\limits_{0}^{\pi}sin(\theta)d\theta\int\limits_{0}^{2\pi}d\phi\ \psi_1^*(r,\theta,\phi)\psi_2(r,\theta,\phi).
\end{equation}
A Galerkin-type approach similar to the one proposed in \cite{Taseli2002}, with spherical harmonics in place of Chebychev basis functions could be considered to solve its eigenvalue problem. However, it is more practical to approach 
the eigenstates perturbationally, starting from the well-known solutions of time-independent \sch equation for  the hydrogenoid atom,
\begin{equation}
\psi^{(0)}_{n,l,m}(r,\theta,\phi)=R_{n,l}(r)Y_{l,m}(\theta,\phi)
\label{H-eigenfunc}
\end{equation}
with
\begin{equation}
R_{n,l}(r)=\left(\frac{2\mu Z}{n}\right)^{\frac{3}{2}}\sqrt{\frac{(n-l-1)!}{2n[(n+l)!]}}exp\left(-\frac{\mu Zr}{n}\right)\left(\frac{2\mu Zr}{n}\right)^l L_{n-l-1}^{2l+1}\left(\frac{2\mu Zr}{n}\right),
\end{equation}
where $L_{n-l-1}^{2l+1}(x)$ denotes the generalized Laguerre polynomials, and $Y_{l,m}(\theta,\phi)$ the spherical harmonics.
We note that the perturbation operator,
\begin{equation}
V^{(1)}(r,\theta):=\left(\frac{1}{a}\right)\frac{\left(1-3 cos(\theta)^2\right) Z}{4 r^3}.
\end{equation}
is proportional to $Y_{2,0}(\theta,\phi)$,
\begin{equation}
V^{(1)}(r,\theta)=\sqrt{\frac{\pi}{5}} \left(\frac{-Z}{a}\right) \frac{Y_{2,0}(\theta,\phi)}{r^3}.
\label{perturb-op}
\end{equation}
Given the following integral formula
\begin{scriptsize}\begin{equation}
\int\limits_{0}^{\pi}sin(\theta)d\theta\int\limits_{0}^{2\pi}d\phi\ Y_{l_1,m_1}(\theta,\phi)Y_{l_2,m_2}(\theta,\phi)Y_{l_3,m_3}(\theta,\phi)=\sqrt{\frac{(2l_1+1)(2l_2+1)(2l_3+1)}{4\pi}}\begin{pmatrix} l_1&l_2&l_3\\
          0&0&0                                                                                                                                                                                                 \end{pmatrix}\begin{pmatrix} l_1&l_2&l_3\\
          m_1&m_2&m_3                                                                                                                                                                                                 \end{pmatrix},
\end{equation}\end{scriptsize}
and the well-known relation for the conjugate of a spherical harmonic:
\begin{equation}
Y^*_{l,m}(\theta,\phi)=(-1)^m Y_{l,-m}(\theta,\phi),
\end{equation}
we deduce that, for a given $(n,l,m)$-triplet of quantum numbers, the state $\psi^{(0)}_{n,l,m}$ can only be coupled at first order to states $\psi^{(0)}_{n',l',m}$'s such that (i) $l'\geq |m|$, (ii) for $l'\in\{|l-2|,\cdots ,l+2\}$, the 3-j symbol\begin{scriptsize} $\begin{pmatrix} l'&l&2\\
          0&0&0                                                                                                                                                                                                 \end{pmatrix}$                                                                                                                                                                                                                              \end{scriptsize}
 is non zero,  and, (iii) $n'>l'$. The allowed quantum number values are summed up in Tab.\ref{couplings}.
 \begin{table}[h!]
\begin{scriptsize}\begin{tabular}{@{}ll}
\hline
Zero-order states& first-order perturbatively coupled states \\
\hline
$n>0, l=0, m=0$ & $n'>2, l'=2, m'=0$ \\
$n>1, l=1, m\in\{-1,0,1\}$ &$n'>1, l'=1, m'=m$, $n'>3, l'=3, m'=m$\\
$n>l\geq 2, m\in\{-l,-l+1,l-1,l\}$ &$n'>l', l'=l, m'=m$, $n'>l'+2, l'=l'+2, m'=m$\\
$n>l\geq 2, m\in\{-l+2,\cdots,l-2\}$ &$n'>l'-2, l'=l-2, m'=m$, $n'>l', l'=l, m'=m$, $n'>l'+2, l'=l'+2, m'=m$\\
\hline
\end{tabular}\end{scriptsize}
\caption{List of hydrogenoid eigenstates coupled by the perturbation operator of Eq.(\ref{perturb-op}) to a given hydrogenoid eigenstate (in terms of their associated quantum numbers).}
\label{couplings}
\end{table}

\textit{A priori}, the perturbation operator needs to be diagonalized first in each degenerate $n$-subspace. The first order correction to the unperturbed energies, (that is the energies of the
hydrogenoid atom, $E^{(0)}_{n,l,m}=- \frac{\mu Z^2}{2n^2}$ in hartree), are the eigenvalues of the matrix, 
\begin{equation}
 (\langle\psi^{(0)}_{n,l,m}|V^{(1)}|\psi^{(0)}_{n,l',m'}\rangle)_{(l,m),(l',m')}=(\delta_{m,m'}\langle\psi^{(0)}_{n,l,m}|V^{(1)}|\psi^{(0)}_{n,l',m}\rangle)_{(l,m),(l',m')}.
\end{equation}
However, for the cases investigated, this matrix is already diagonal, due to the cancellation of the radial integral.
So, for $l=0$, there is no first order correction,
\begin{eqnarray}
&E_{n,0,0}=- \frac{\mu Z^2}{2n^2},
\label{1ord-eigenval-l=0}
\end{eqnarray}
and for $l>0$,  the correction is,
\begin{eqnarray}
E_{n,l,m}&=&E^{(0)}_{n,l,m}-\sqrt{\frac{\pi}{5}}\left(\frac{Z}{a}\right)\langle\psi^{(0)}_{n,l,m}|\frac{Y_{2,0}(\theta,\phi)}{r^3}|\psi^{(0)}_{n,l,m}\rangle\nonumber\\
&=&- \frac{\mu Z^2}{2n^2}-\sqrt{\frac{\pi}{5}}\left(\frac{Z}{a}\right)\int\limits_{0}^{+\infty}\frac{R^2_{n,l}(r)}{r} dr\times(-1)^m(l+\frac{1}{2})\sqrt{\frac{5}{\pi}}\mbox{\begin{scriptsize}$\begin{pmatrix} l&l&2\\
          0&0&0                                                                                                                                                                                                 \end{pmatrix}\begin{pmatrix} l&l&2\\
          -m&m&0                                                                                                                                                                                                 \end{pmatrix} $                                                                                                                                                                                                                             \end{scriptsize}} \nonumber\\
&=&- \frac{\mu Z^2}{2n^2}-\left(\frac{Z}{a}\right)\times(-1)^m(l+\frac{1}{2})\mbox{\begin{scriptsize}$\begin{pmatrix} l&l&2\\
          0&0&0                                                                                                                                                                                                 \end{pmatrix}\begin{pmatrix} l&l&2\\
          -m&m&0                                                                                                                                                                                                 \end{pmatrix} $                                                                                                                                                                                                                             \end{scriptsize}}\int\limits_{0}^{+\infty}\frac{R^2_{n,l}(r)}{r} dr .
\label{1ord-eigenval-l>0}
\end{eqnarray}
As expected, the spherical symmetry is broken:  an $(l,|m|)$-dependency is introduced in the perturbed eigenvalues. 
In Tab.\ref{energies}, we provide the first order eigenvalues, which can be useful for basis set truncation purposes.
We note that, at first order, degeneracy is not completely lifted, as for example,  $E_{4,0,0}=E_{4,3,\pm 2}$. 
If we factorize by the zero order energy, we see that the relative corrections at the first order are all proportional to $\frac{\mu^2 Z^2}{a}$.\\ 
A similar observation can be made for the first order corrected eigenfunctions,
\begin{eqnarray}
\psi_{n,l,m}&=&\psi^{(0)}_{n,l,m}+\sum\limits_{(n',l')\neq (n,l)}\frac{-\sqrt{\frac{\pi}{5}}\left(\frac{Z}{a}\right)\langle\psi^{(0)}_{n',l',m}|\frac{Y_{2,0}(\theta,\phi)}{r^3}|\psi^{(0)}_{n,l,m}\rangle}{E^{(0)}_{n,l,m}-E^{(0)}_{n',l',m}}\ \psi^{(0)}_{n',l',m}\nonumber\\
&=&\psi^{(0)}_{n,l,m}+\sum\limits_{(n',l')\neq (n,l)}\frac{-\left(\frac{Z}{a}\right)\int\limits_{0}^{+\infty}\frac{R_{n',l'}(r)R_{n,l}(r)}{r} dr\times\mbox{\begin{scriptsize}$(-1)^m\sqrt{(l'+\frac{1}{2})(l+\frac{1}{2})}\begin{pmatrix} l'&l&2\\
          0&0&0                                                                                                                                                                                                 \end{pmatrix}\begin{pmatrix} l'&l&2\\
          -m&m&0                                                                                                                                                                                                 \end{pmatrix} $                                                                                                                                                                                                                             \end{scriptsize}} }{- \frac{\mu Z^2}{2n^2}+ \frac{\mu Z^2}{2n'^2}}\psi^{(0)}_{n',l',m}\nonumber\\
&=&\psi^{(0)}_{n,l,m}+\sum\limits_{(n',l')\neq (n,l)}\mbox{\begin{scriptsize}$\frac{2n'^2n^2}{a\mu Z(n'^2-n^2)}(-1)^m\sqrt{(l'+\frac{1}{2})(l+\frac{1}{2})}\begin{pmatrix} l'&l&2\\
          0&0&0                                                                                                                                                                                                 \end{pmatrix}\begin{pmatrix} l'&l&2\\
          -m&m&0                                                                                                                                                                                                 \end{pmatrix}\int\limits_{0}^{+\infty}\frac{R_{n',l'}(r)R_{n,l}(r)}{r} dr $                                                                                                                                                                                                                             \end{scriptsize}} \psi^{(0)}_{n',l',m},\nonumber\\
\label{1ord-eigenstate}
\end{eqnarray}
the radial integral being proportional to $\mu^3 Z^3$ (because of the $\frac{1}{r^3}$-term, when we make the change of variables $x=2\mu Z r$), 
all the coupling coefficients in the expansion are proportional to $\frac{\mu^2 Z^2}{a}$. 
Here also, the potentially divergent terms in the expansion, due to the degeneracy of the zero$^{th}$ order eigenvalues for a given $n$ in the denominators, can be excluded since the corresponding numerators cancel out because of the radial integrals again.\\

The first terms in the expansion of the lowest eigenfunctions are given in Tab. \ref{first-eigenfunc}.

 \begin{table}[ht!]
\begin{scriptsize}
\begin{tabular}{l|llll}
\hline
&$l=0$&$l=1$&$l=2$&$l=3$\\
\hline
\multicolumn{5}{c}{}\\[20pt]
&&$E_{4,1,\pm 1}=- \frac{\mu Z^2}{32} + \frac{\mu^3 Z^4}{1920 a}$&&\\[5pt]
&&&$E_{4,2,\pm 2}=- \frac{\mu Z^2}{32}+  \frac{\mu^3 Z^4}{6760 a}$&\\[5pt]
&&&&$E_{4,3,\pm 3}=- \frac{\mu Z^2}{32}+  \frac{\mu^3 Z^4}{16128 a}$\\[5pt]
$n=4$&$E_{4,0,0}=- \frac{\mu Z^2}{32}$ &&&$E_{4,3,\pm 2}=- \frac{\mu Z^2}{32}$\\[5pt]
&&&&$E_{4,3,\pm 1}=- \frac{\mu Z^2}{32}-  \frac{\mu^3 Z^4}{26880 a}$\\[5pt]
&&&&$E_{4,3,0}=- \frac{\mu Z^2}{32}-  \frac{\mu^3 Z^4}{20160 a}$\\[5pt]
&&&$E_{4,2,\pm 1}=- \frac{\mu Z^2}{32} - \frac{\mu^3 Z^4}{13440 a}$&\\[5pt]
&&&$E_{4,2,0}=- \frac{\mu Z^2}{32} - \frac{\mu^3 Z^4}{6760 a}$&\\[5pt]
& &$E_{4,1,0}=- \frac{\mu Z^2}{32} - \frac{\mu^3 Z^4}{960 a}$&&\\[20pt]
\multicolumn{5}{c}{....................................................................................................................}\\[20pt]
&&$E_{3,1,\pm 1}=- \frac{\mu Z^2}{18} + \frac{\mu^3 Z^4}{810 a}$&&\\[5pt]

&&&$E_{3,2,\pm 2}=- \frac{\mu Z^2}{18}+  \frac{\mu^3 Z^4}{2835 a}$&\\[5pt]
$n=3$&$E_{3,0,0}=- \frac{\mu Z^2}{18}$ &&&\\[5pt]
&&&$E_{3,2,\pm 1}=- \frac{\mu Z^2}{18} - \frac{\mu^3 Z^4}{5670 a}$&\\[5pt]
&&&$E_{3,2,0}=- \frac{\mu Z^2}{18} - \frac{\mu^3 Z^4}{2835 a}$&\\[5pt]
& &$E_{3,1,0}=- \frac{\mu Z^2}{18} - \frac{\mu^3 Z^4}{405 a}$&&\\[20pt]
\multicolumn{5}{c}{....................................................................................................................}\\[20pt]
&&$E_{2,1,\pm 1}=- \frac{\mu Z^2}{8} + \frac{\mu^3 Z^4}{240 a}$&&\\[5pt]
$n=2$&$E_{2,0,0}=- \frac{\mu Z^2}{8}$ &&&\\[5pt]
&&$E_{2,1,0}=- \frac{\mu Z^2}{8} - \frac{\mu^3 Z^4}{120 a}$&&\\[20pt]
\multicolumn{5}{c}{....................................................................................................................}\\[20pt]
$n=1$&$E_{1,0,0}=- \frac{\mu Z^2}{2}$ &&&\\[5pt]

\hline
\end{tabular}\end{scriptsize}
\caption{First-order corrected energies (up to n=4). For every pairs, $(n,l)$, the sum over $m\in\{-l,-l+1,\cdots,l-1,l\}$ of the first order corrections is zero. The spacing between the energies does not follow any scale, only the order between the levels is respected. }
\label{energies}
\end{table}

\newpage

\begin{table}[ht!]
\begin{scriptsize}
\begin{tabular}{lll|l}
\hline
$n=1$&$l=0$&$m=0$&$\psi_{1,0,0}=\psi^{(0)}_{1,0,0}+\frac{\sqrt{5}}{5 a\mu Z} \sum\limits_{n'>2}\mbox{\begin{scriptsize}$\frac{1}{ 1-\frac{1}{n'^2}}\int\limits_{0}^{+\infty}\frac{R_{n',2}(r)R_{1,0}(r)}{r} dr $ \end{scriptsize}} \psi^{(0)}_{n',2,0}$\\  
\multicolumn{3}{c|}{} & \hspace{0.9cm}$=\psi^{(0)}_{1,0,0}+\frac{\mu^2 Z^2}{a}\left(\frac{\sqrt{6}}{480 }\psi^{(0)}_{3,2,0}+\frac{104}{28125}  \psi^{(0)}_{4,2,0}+\frac{5\sqrt{14}}{6804} \psi^{(0)}_{5,2,0}+\frac{5744\sqrt{21}}{12353145}\psi^{(0)}_{6,2,0}+\frac{3299 \sqrt{21} }{8847360}\psi^{(0)}_{7,2,0}+\cdots\right)$\\ 
$n=2$&$l=1$&$m=0$&$\psi_{2,1,0}=\psi^{(0)}_{2,1,0}+\frac{8}{5 a\mu Z} \sum\limits_{n'>2}\mbox{\begin{scriptsize}$\frac{1}{ 1-\frac{4}{n'^2}}\int\limits_{0}^{+\infty}\frac{R_{n',1}(r)R_{2,1}(r)}{r} dr $ \end{scriptsize}} \psi^{(0)}_{n',1,0}+\frac{12\sqrt{21}}{35 a\mu Z} \sum\limits_{n'>3}\mbox{\begin{scriptsize}$\frac{1}{ 1-\frac{4}{n'^2}}\int\limits_{0}^{+\infty}\frac{R_{n',3}(r)R_{2,1}(r)}{r} dr $ \end{scriptsize}} \psi^{(0)}_{n',3,0}$\\  
\multicolumn{3}{c|}{} &\hspace{0.8cm} $=\psi^{(0)}_{2,1,0}+\frac{\mu^2 Z^2}{a}\left(\frac{192}{3125}\psi^{(0)}_{3,1,0}+\frac{56\sqrt{10}}{6075}\psi^{(0)}_{4,1,0}+\frac{8768\sqrt{5}}{1058841}  \psi^{(0)}_{5,1,0}+\frac{201 \sqrt{35}}{89600} \psi^{(0)}_{6,1,0}+\frac{2926784\sqrt{14}}{1076168025}\psi^{(0)}_{7,1,0}+\cdots\right)$\\ 
\multicolumn{3}{c|}{} & \hspace{1.9cm} $+\frac{\mu^2 Z^2}{a}\left(\frac{8 \sqrt{10}}{14175}\psi^{(0)}_{4,3,0}+\frac{512 \sqrt{5}}{823543}  \psi^{(0)}_{5,3,0}+\frac{123 \sqrt{10}}{358400} \psi^{(0)}_{6,3,0}+\frac{541184\sqrt{3}}{1076168025}\psi^{(0)}_{7,3,0}+\cdots\right)$\\ 
$n=2$&$l=0$&$m=0$&$\psi_{2,0,0}=\psi^{(0)}_{2,0,0}+\frac{4\sqrt{5}}{5 a\mu Z} \sum\limits_{n'>2}\mbox{\begin{scriptsize}$\frac{1}{ 1-\frac{4}{n'^2}}\int\limits_{0}^{+\infty}\frac{R_{n',2}(r)R_{2,0}(r)}{r} dr $ \end{scriptsize}} \psi^{(0)}_{n',2,0}$\\  
\multicolumn{3}{c|}{} & \hspace{0.9cm}$=\psi^{(0)}_{2,0,0}-\frac{\mu^2 Z^2}{a}\left(\frac{32 \sqrt{3}}{9375}\psi^{(0)}_{3,2,0}+ \frac{2 \sqrt{2}}{1215}  \psi^{(0)}_{4,2,0}+\frac{3680\sqrt{7}}{7411887} \psi^{(0)}_{5,2,0}+\frac{29\sqrt{42}}{215040}\psi^{(0)}_{6,2,0}+\frac{63584 \sqrt{42} }{645700815}\psi^{(0)}_{7,2,0}+\cdots\right)$\\ 
$n=2$&$l=1$&$m=\pm 1$&$\psi_{2,1,\pm 1}=\psi^{(0)}_{2,1,\pm 1}-\frac{4}{5 a\mu Z} \sum\limits_{n'>2}\mbox{\begin{scriptsize}$\frac{1}{ 1-\frac{4}{n'^2}}\int\limits_{0}^{+\infty}\frac{R_{n',1}(r)R_{2,1}(r)}{r} dr $ \end{scriptsize}} \psi^{(0)}_{n',1,\pm 1}$\\
\multicolumn{3}{c|}{} &\hspace{2.3cm} $+\frac{12\sqrt{14}}{35 a\mu Z} \sum\limits_{n'>3}\mbox{\begin{scriptsize}$\frac{1}{ 1-\frac{4}{n'^2}}\int\limits_{0}^{+\infty}\frac{R_{n',3}(r)R_{2,1}(r)}{r} dr $ \end{scriptsize}} \psi^{(0)}_{n',3,\pm 1}$\\
\multicolumn{3}{c|}{} & $=\psi^{(0)}_{2,1,\pm 1}-\frac{\mu^2 Z^2}{a}\left(\frac{96}{3125}\psi^{(0)}_{3,1,\pm 1}+\frac{28 \sqrt{10}}{18225}\psi^{(0)}_{4,1,\pm 1}+\frac{4384 \sqrt{5}}{1058841}  \psi^{(0)}_{5,1,\pm 1}+\frac{201 \sqrt{35}}{179200} \psi^{(0)}_{6,1,\pm 1}+\frac{1463392\sqrt{14}}{1076168025 }\psi^{(0)}_{7,1,\pm 1}+\cdots\right)$\\
\multicolumn{3}{c|}{} & \hspace{0.06cm}$ \qquad  \qquad +\frac{\mu^2 Z^2}{a}\left(\frac{16 \sqrt{15}}{42525}\psi^{(0)}_{4,3,\pm 1}+\frac{512 \sqrt{30}}{2470629}  \psi^{(0)}_{5,3,\pm 1}+\frac{41 \sqrt{15}}{179200} \psi^{(0)}_{6,3,\pm 1}+\frac{541184\sqrt{2}}{1076168025 }\psi^{(0)}_{7,3,\pm 1}+\cdots\right)$\\
$n=3$&$l=1$&$m=0$&$\psi_{3,1,0}=\psi^{(0)}_{3,1,0}+\frac{18}{5 a\mu Z} \sum\limits_{\substack{n'>1 \\ n'\neq 3}}\mbox{\begin{scriptsize}$\frac{1}{ 1-\frac{9}{n'^2}}\int\limits_{0}^{+\infty}\frac{R_{n',1}(r)R_{3,1}(r)}{r} dr $ \end{scriptsize}} \psi^{(0)}_{n',1,0}+\frac{27\sqrt{21}}{35 a\mu Z} \sum\limits_{n'>3}\mbox{\begin{scriptsize}$\frac{1}{ 1-\frac{9}{n'^2}}\int\limits_{0}^{+\infty}\frac{R_{n',3}(r)R_{3,1}(r)}{r} dr $ \end{scriptsize}} \psi^{(0)}_{n',3,0}$\\  
\multicolumn{3}{c|}{} & $=\psi^{(0)}_{3,1,0}+\frac{\mu^2 Z^2}{a}\left(\frac{-192}{3125}\psi^{(0)}_{2,1,0}+\frac{60288\sqrt{10}}{2941225}\psi^{(0)}_{4,1,0}+\frac{231\sqrt{5}}{16384}  \psi^{(0)}_{5,1,0}+\frac{35648 \sqrt{35}}{10333575} \psi^{(0)}_{6,1,0}+\frac{310191\sqrt{14}}{78125000}\psi^{(0)}_{7,1,0}+\cdots\right)$\\ 
\multicolumn{3}{c|}{} & \hspace{1.9cm} $-\frac{\mu^2 Z^2}{a}\left(\frac{10368 \sqrt{10}}{20588575}\psi^{(0)}_{4,3,0}+\frac{81 \sqrt{5}}{458752}  \psi^{(0)}_{5,3,0}+\frac{128 \sqrt{10}}{3444525} \psi^{(0)}_{6,3,0}+\frac{567
\sqrt{3}}{39062500}\psi^{(0)}_{7,3,0}+\cdots\right)$\\ 
\hline
\end{tabular}
\end{scriptsize}
\caption{First-order corrected wave functions, ordered in increasing energy eigenvalue up to n=4 (in Appendix we provide a more comprehensive table up to n=7, ``i-orbitals'').}
\label{first-eigenfunc}
\end{table}

\newpage

\begin{center}

\begin{figure}[h!]

\begin{tabular}{cccc}
&$1s$&$2p_z$&$2p_x$\\[00pt]
\hline 
\raisebox{40pt}{
$V(\vec{r})=\frac{-Z}{||\vec{r}||}$}&\includegraphics*[scale=0.3]{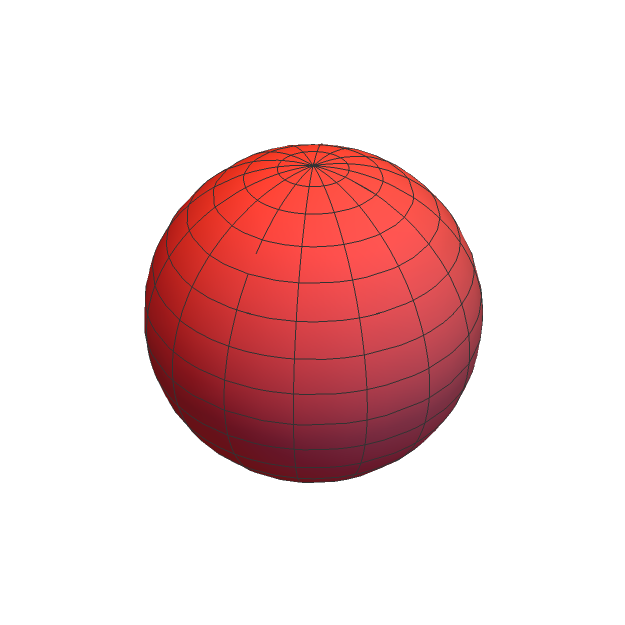}&\includegraphics*[scale=0.3]{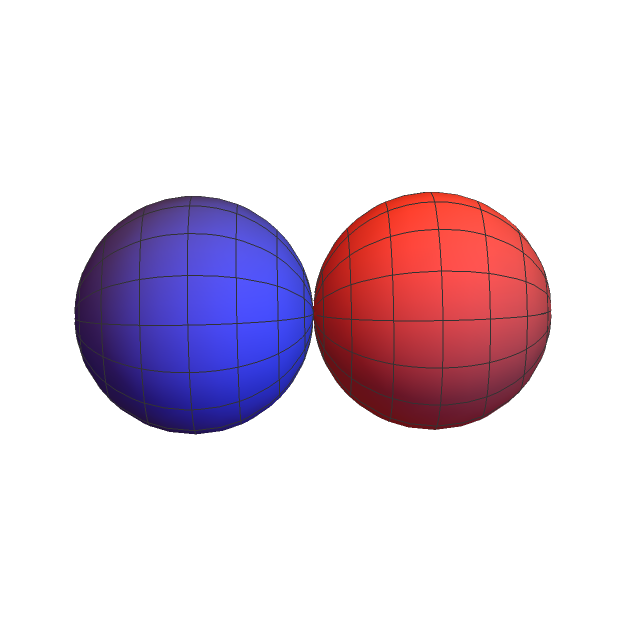}&\includegraphics*[scale=0.3]{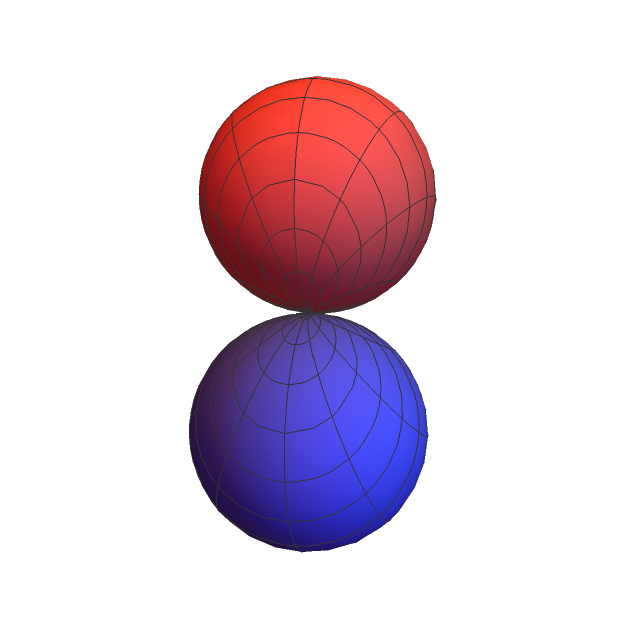}\\[00pt]
\raisebox{40pt}{$V(\vec{r})=-Z\sqrt{\frac{a}{\pi}}\int\limits_{-\infty}^{+\infty} \frac{exp[-az_0^2]}{||\vec{r}-\vec{r_{z_0}}||} dz_0$}&\includegraphics*[scale=0.3]{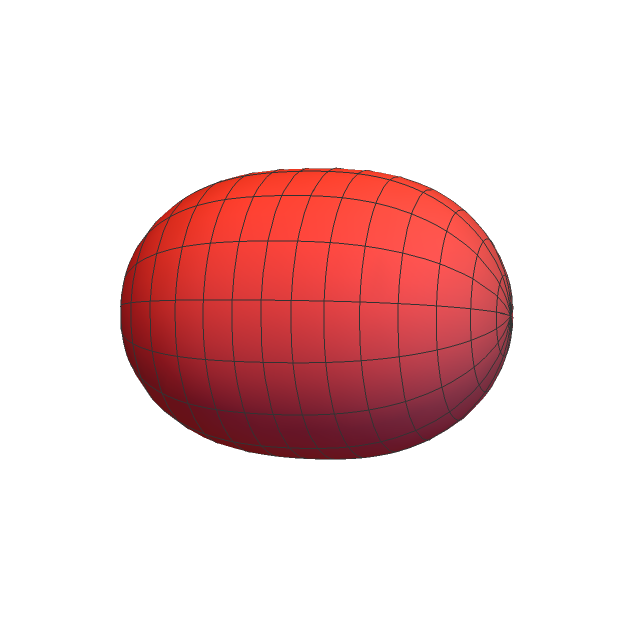}&\includegraphics*[scale=0.3]{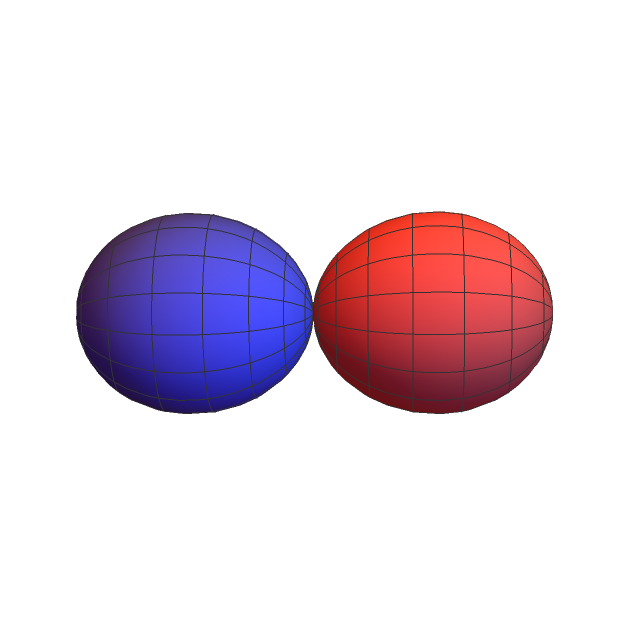}&\includegraphics*[scale=0.3]{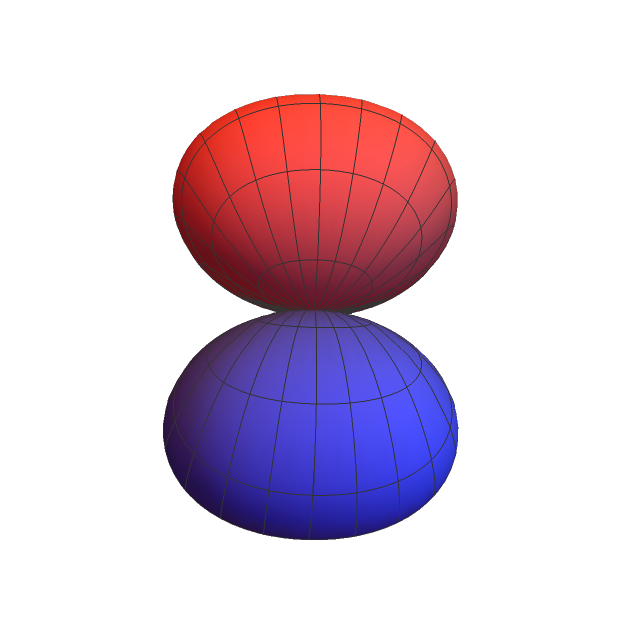}\\
\hline             
\end{tabular}

\caption{Comparison of hydrogenoid and smeared Coulomb potential orbitals. The orbitals represented in the second row correspond are the first order correction for a first order expansion of the smeared Coulomb potential in $\frac{1}{a}$ as $a\rightarrow+\infty$. The parameters are tuned so that the corrective contributions are about $10\%$ of the unperturbed solutions displayed in the first row. 
\label{orbitals-sp}}

\end{figure}
\end{center}

\section{Conclusion}

In this study, we have obtained first order perturbative corrections to the eigenfunctions of a smeared Coulomb potential along an arbitrary axis. Such a potential can represent the potential felt by an electron bounded to a vibrating nucleus of effective charge $Z$, effective reduced mass $\mu$ and effective classical vibrating amplitude equal to $\sqrt{\frac{2n_{vib}+1}{2a}}$ (in harmonic quantum level $ n_{vib}$). Hence, it is hoped that these approximate eigenstates can be appropriate to describe the electron density of the effective electronic Hamiltonians solved in the EN-MFCI method. On the practical side, it can be taken advantage of the remarkable fact that,  only a single parameter, namely $\frac{\mu^2Z^2}{a}$, need to be optimized for all the basis functions. 

Furthermore, the DHO basis functions can be seen as contracted STO, corresponding to different angular momenta. So, the  codes and techniques already developed for STO basis functions, can conveniently be employed to compute the multicenter integrals, that are needed for molecular computations with these new basis functions.

A hyperbolic cosine factor can be associated to ETO to provide a ``double zeta'' character to a minimal basis set \cite{Sahin2017}. This can be considered for DHO, as well. However, one can also combine different sets of DHO corresponding to atoms at different ionization states, for example, to obtain proper multi-zeta basis sets.

Beside the application to the EN-MFCI method, these orbitals could be useful for ``clamped nuclei'' quantum chemistry calculations, as they can be regarded as naturally ``hybridized'' (unlike spherical harmonics STO). One can develop a battery of model Hamiltonian adapted to one or several bond directions (whatever a chemical bond might be from the quantum mechanical point of view) and obtain their first order approximate eigenfunctions by following the approach presented in this paper. For example, different shapes of Coulomb potentials could provide different model Hamitonians for the sp, sp2, or sp3 hybridization of a carbon atom,  and in turn, relevant naturally-hybridized orbital eigenfunction basis sets.

Another potential application  for ``clamped nuclei'' quantum chemistry is to save on polarization orbital sets, since high angular momentum atomic primitives are already contracted within our orbitals.
%

\end{document}